\documentclass[]{aa}

\def\rmit#1{{\it #1}}              

\def\eg{\rmit{e.g.}}

\usepackage{natbib}
\bibpunct{(}{)}{;}{a}{}{,} 
\usepackage{graphicx}
\usepackage{amsmath}

\usepackage{color}

\usepackage{hyperref}
\hypersetup{
    final=true,
    pageanchor=true,
    colorlinks=true,
    breaklinks=true,
    linkcolor=blue,
    citecolor=blue,
    urlcolor=blue,
    pdfpagemode=UseNone,
    pdftitle={Helioseismic Holography of Simulated Sunspots},
    pdfauthor={T. Felipe et al.},
    pdfsubject={Solar Physics},
    pdfkeywords={methods: numerical - Sun: helioseismology - Sun: oscillations - sunspots - Sun: interior - Sun: magnetic fields}}

\titlerunning{Helioseismic Holography of Simulated Sunspots}

\authorrunning{Felipe et al.}

\begin{document}

\title{Helioseismic Holography of Simulated Sunspots: dependence of the travel time on magnetic field strength and Wilson depression}

\author{T. Felipe\inst{\ref{inst1},\ref{inst2}}
\and D. C. Braun\inst{\ref{inst3}}
\and A. C. Birch\inst{\ref{inst4}}
}


\institute{Instituto de Astrof\'{\i}sica de Canarias, 38205, C/ V\'{\i}a L{\'a}ctea, s/n, La Laguna, Tenerife, Spain\label{inst1}
\and 
Departamento de Astrof\'{\i}sica, Universidad de La Laguna, 38205, La Laguna, Tenerife, Spain\label{inst2}
\and 
NorthWest Research Associates, Boulder, CO 80301, USA\label{inst3} 
\and 
Max-Planck-Institut f\"{u}r Sonnensystemforschung, Justus-von-Liebig-Weg 3, 37077 G\"{o}ttingen, Germany\label{inst4} 
}

\abstract{Improving methods for determining the subsurface structure of sunspots from their seismic signature requires a better understanding of the interaction of waves with magnetic field concentrations. We aim to quantify the impact of changes in the internal structure of sunspots on local helioseismic signals. We have numerically simulated the propagation of a stochastic wave field through sunspot models with different properties, accounting for changes in the Wilson depression between 250 and 550 km and in the photospheric umbral magnetic field between 1500 and 3500 G. The results show that travel-time shifts at frequencies above approximately 3.50 mHz (depending on the phase-speed filter) are insensitive to the magnetic field strength. The travel time of these waves is determined exclusively by the Wilson depression and sound-speed perturbation. The travel time of waves with lower frequencies is affected by the direct effect of the magnetic field, although photospheric field strengths below 1500 G do not leave a significant trace on the travel-time measurements. These results could potentially be used to develop simplified travel-time inversion methods.}


\maketitle


\section{Introduction}

Solar active regions, such as sunspots, are the most remarkable manifestations of solar magnetism and have a key role in the coupling between the interior and the atmosphere, which makes the understanding of sunspot formation, stability, and decay one of the outstanding problems in solar physics. Local helioseismic studies have been focused on solar active regions for more than two decades, but still significant uncertainty exists regarding their interpretation due to the complexity of wave interactions with magnetic fields.

Several helioseismic techniques have been developed to probe the subsurface structure of active regions. The most commonly used measurement is the phase change of waves traveling through the active region, usually described as a change in travel time, which can be obtained from time-distance helioseismology \citep{Duvall+etal1993}, Fourier-Hankel analysis \citep{Braun+etal1992}, or helioseismic holography \citep{Braun+Birch2008}. The first attempts to interpret the seismic signals assumed that sunspots can be characterized as small perturbations to the thermal structure of a quiet Sun model and neglected the direct effects of the magnetic field on helioseismic waves. From this approach two different sunspot scenarios were inferred from observations: a shallow positive perturbation of the sound speed \citep{Fan+etal1995} and a two-layer model with a reduction of the wave-speed in the top layer and an increase in the wave-speed down to 10 Mm below the surface \citep{Kosovichev1996, Kosovichev+etal2000}.       

Subsequent observational \citep{Lindsey+Braun2005, Schunker+etal2005, Couvidat+Rajaguru2007} and numerical \citep{Cameron+etal2008, Cally2009, Moradi+etal2009, Cameron+etal2011, Moradi+etal2015} studies have shown that surface magnetic fields strongly alter travel times. In addition to the changes in the phase speed of the waves introduced by the magnetic field, mode conversion \citep{Schunker+Cally2006, Cally+Goossens2008} can also affect the photospheric seismic measurements through the returning of atmospheric fast waves \citep{Rosenthal+etal2002, Khomenko+Collados2006, Felipe+etal2010a} or Alfv\'en waves \citep{Cally+Hansen2011, Hansen+Cally2012, Khomenko+Cally2012, Felipe2012}. \citet{Cally+Moradi2013} quantified the contribution of these returned waves to the travel-time shifts obtained from the analysis of the photospheric wave field. They found a significant travel-time shift which depends on the propagation direction of the waves and the magnetic field inclination. 

Recently, \citet{Felipe+etal2016a} evaluated the individual contributions of the direct and indirect magnetic effects on the travel-time perturbations measured using helioseismic holography. The term ``direct magnetic effects'' refers to the changes in wave propagation due to the wave interaction with the magnetic field through mode conversion and fast wave refraction, which produces variations in the wave speed and ray paths. The ``indirect magnetic effects'' account for the changes in the thermal structure of the sunspot imposed by the presence of the magnetic field. Their results show that waves filtered for certain phase speeds and frequencies are less sensitive to the magnetic field. For those combinations of filters, the travel-time shifts are accurately predicted from the thermal structure of the sunspot, accounting for the Wilson depression and the changes in the sound speed but neglecting the direct effect of the magnetic field. In this work, we aim to further explore the dependence of the helioseismic signals on sunspot properties by analyzing a larger sample of sunspot models which differ in their magnetic field strength and Wilson depression. The numerical methods are described in \hbox{Sect. \ref{sect:simulations}} and the comparison of the results obtained for different models is presented in \hbox{Sect. \ref{sect:results}}. Finally, the conclusions and the potential applications of our results are discussed in \hbox{Sect. \ref{sect:conclusions}}.

\begin{figure}[!ht] 
 \centering
 \includegraphics[width=9cm]{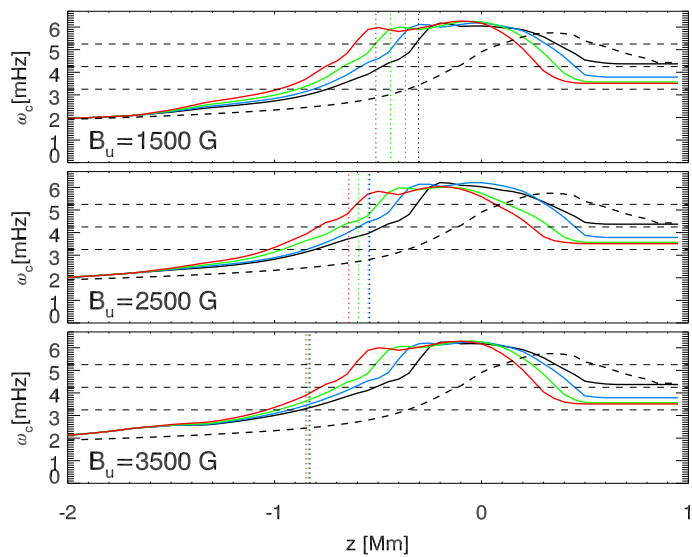}
  \caption{Vertical stratification of the isothermal cut-off frequency at the axis of the sunspot models with $B_{\rm u}=1500$ G (top panel), $B_{\rm u}=2500$ G (middle panel), and $B_{\rm u}=3500$ G (bottom panel), where $z$ is height and $z=0$ is defined at the quiet-Sun photosphere. The colored lines correspond to the Wilson depression of the model: $z_{\rm WD}=250$ km (black), $z_{\rm WD}=350$ km (blue), $z_{\rm WD}=450$ km (green), and $z_{\rm WD}=550$ km (red). The vertical dotted lines illustrate the height where $\beta=1$ for the model with the Wilson depression associated to the corresponding color. The horizontal thin dashed lines mark some selected frequencies (3.25, 4.25, 5.25 mHz). The thick black dashed line illustrates the cut-off frequency at the quiet-Sun.}
  \label{fig:cutoffs}
\end{figure}

\begin{figure}[!ht] 
 \centering
 \includegraphics[width=9cm]{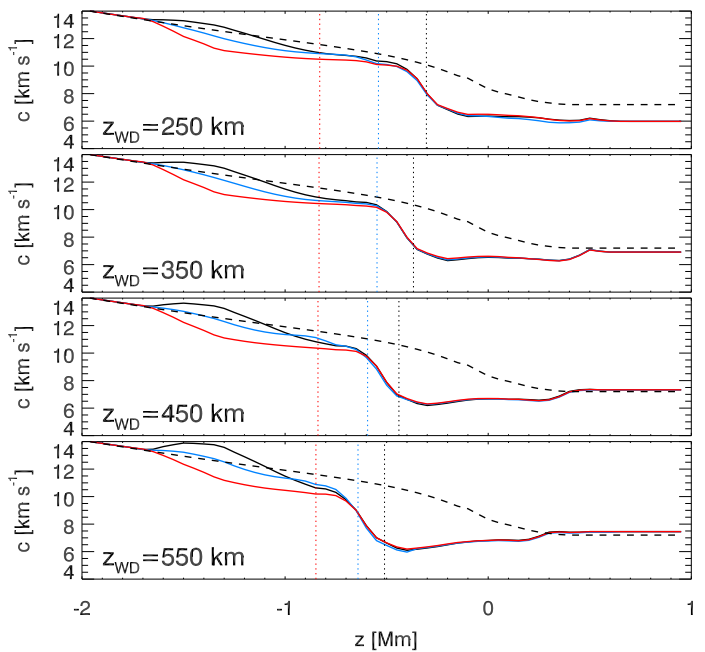}
  \caption{Vertical stratification of the sound speed at the axis of the sunspot models with $z_{\rm WD}=250$ km (top panel), $z_{\rm WD}=350$ km (second panel), $z_{\rm WD}=450$ km (third panel), and $z_{\rm WD}=550$ km (bottom panel), where $z$ is height and $z=0$ is defined at the quiet-Sun photosphere. The color of the lines indicates the photospheric umbral magnetic field strength of the model: $B_{\rm u}=1500$ G km (black), $B_{\rm u}=2500$ G (blue), and $B_{\rm u}=3500$ G km (red). The vertical dotted lines illustrate the height where $\beta=1$ for the model with the magnetic field associated to the corresponding color. The thick black dashed line illustrates the sound speed in the quiet Sun.}
  \label{fig:csonido}
\end{figure}

\begin{figure}[!ht] 
 \centering
 \includegraphics[width=9cm]{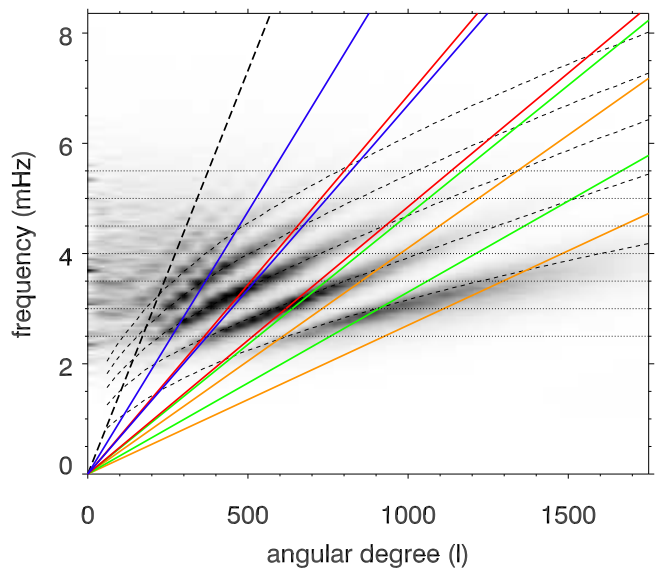}
  \caption{Power spectrum of the vertical velocity at $\log\tau=0.01$ from the quiet Sun simulation. The gray scale represents the power from low (white) to high (black). The thin dashed lines indicate the position of the eigenfrequencies of the $f$ and $p-$modes in model S. The thick dashed line illustrates the region where the phase speed is equal to the sound speed at the bottom boundary of the model. Regions between lines with the same color correspond to the range in phase speed spanned by phase-speed filters TD2 (orange), TD3 (green), TD4 (red), and TD5 (dark blue). The horizontal dotted lines at constant frequency mark the bandpass regions of the frequency filters.}
  \label{fig:power_spectra}
\end{figure}

\begin{figure}[!ht] 
 \centering
 \includegraphics[width=9cm]{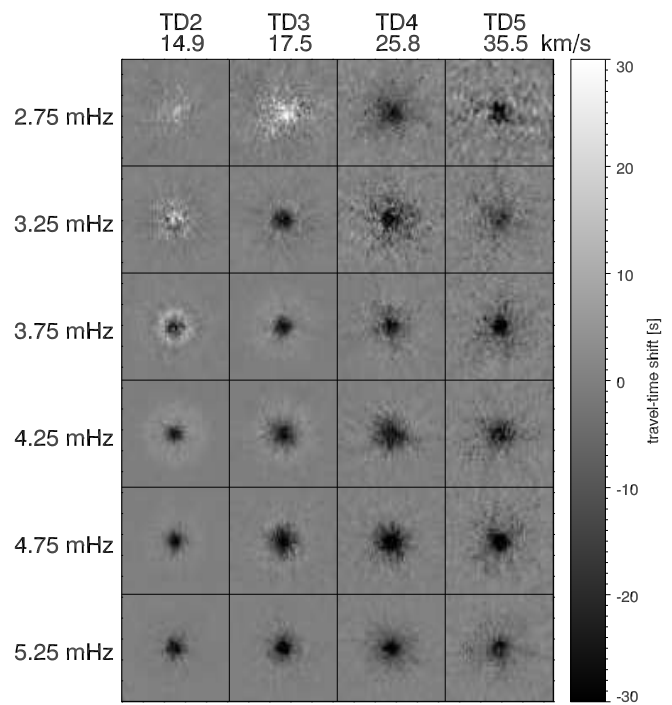}
  \caption{Maps of travel-time shifts obtained for the sunspot model with $B_{\rm u}=1500$ G and $z_{\rm WD}=550$ km using phase speed and frequency bandpass filters. Each column corresponds to a different phase-speed filter (the name of the filter and its phase speed in km s$^{-1}$ are indicated at the top). The rows correspond to frequency filters (increasing from 2.75 mHz in the top row to 5.25 mHz in the bottom row).}
  \label{fig:maps B1500G WD550km}
\end{figure}

\begin{figure}[!ht] 
 \centering
 \includegraphics[width=9cm]{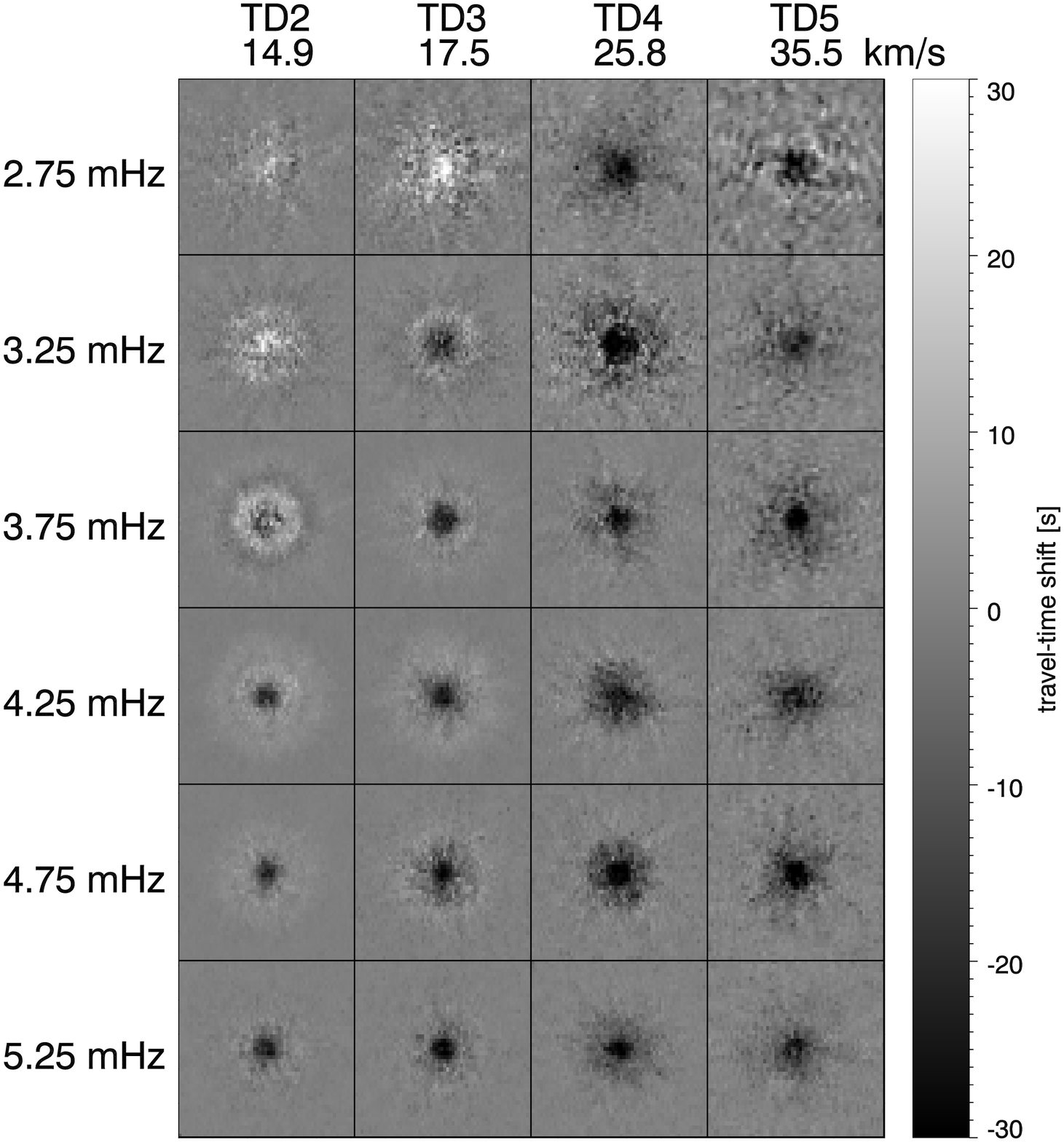}
  \caption{Same as Fig. \ref{fig:maps B1500G WD550km} but for the sunspot with $B_{\rm u}=2500$ G and $z_{\rm WD}=550$ km.}
  \label{fig:maps B2500G WD550km}
\end{figure}

\begin{figure}[!ht] 
 \centering
 \includegraphics[width=9cm]{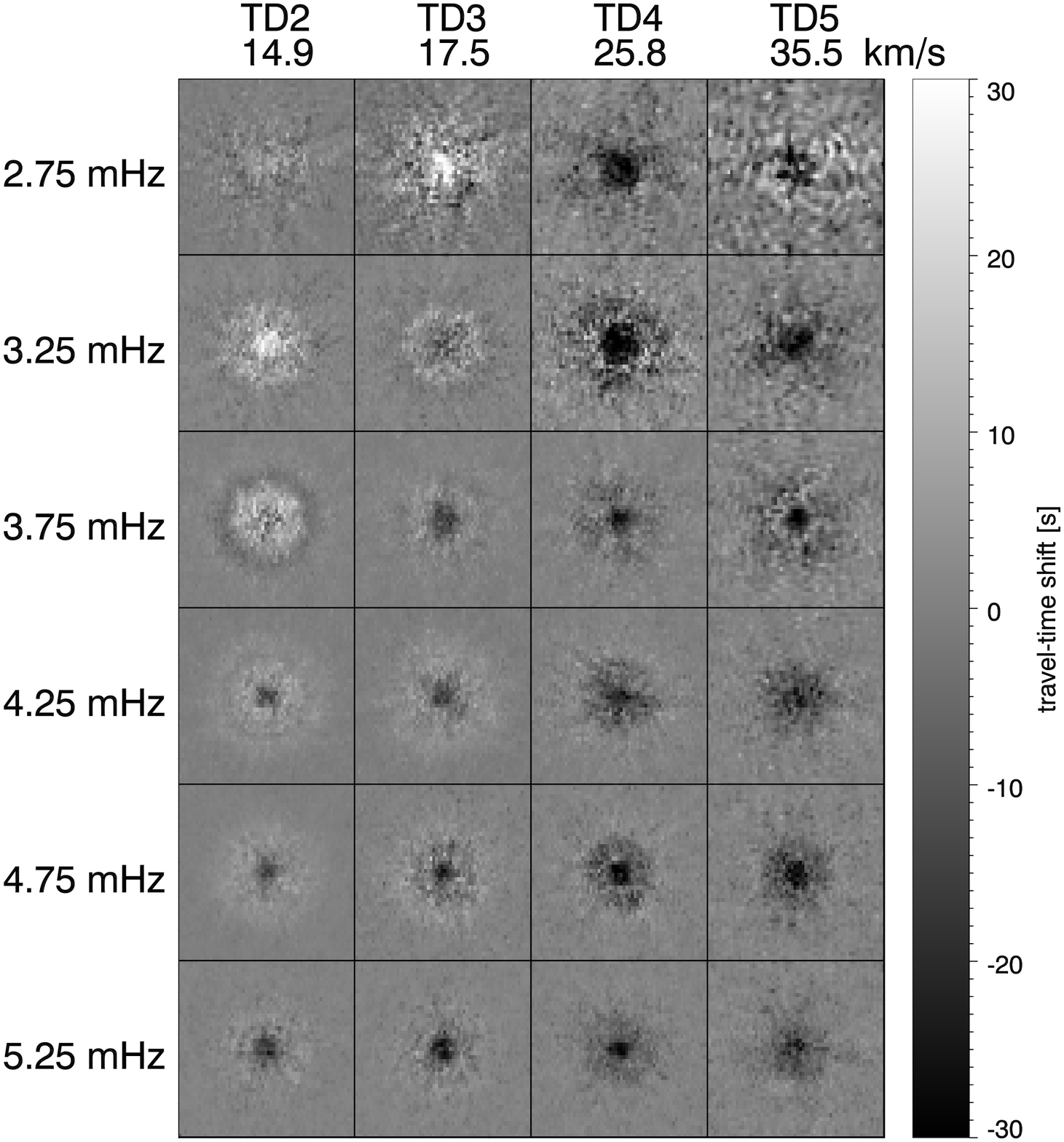}
  \caption{Same as Fig. \ref{fig:maps B1500G WD550km} but for the sunspot with $B_{\rm u}=3500$ G and $z_{\rm WD}=550$ km.}
  \label{fig:maps B3500G WD550km}
\end{figure}

\section{Numerical procedures}
\label{sect:simulations}

We carried out twelve numerical simulations of wave propagation through magnetohydrostatic (MHS) sunspot models using the code MANCHA \citep{Khomenko+Collados2006, Felipe+etal2010a}. The numerical set up of all the simulations is the same described in \citet{Felipe+etal2016a}. Here we include only a brief description of the numerical configuration. We address the interested reader to \citet{Felipe+etal2016a} for a detailed discussion of the numerical methods. The vertical computational domain spans from \hbox{$z=-25$ Mm} to \hbox{$z=1$ Mm}, with $z=0$ defined at the quiet-Sun photosphere and a uniform vertical grid size of 50 km. The dimension of the horizontal directions is 102.4 Mm and they are sampled with a resolution of 200 km. The axes of the sunspots are located at the middle of the domain. The Alfv\'en speed has been artificially limited to \hbox{80 km s$^{-1}$} following \citet{Rempel+etal2009}.   

The simulations differ in the properties of the MHS models used as background. All the sunspot models were constructed with the same method \citep{Khomenko+Collados2008}, but changing two of their properties: magnetic field strength and Wilson depression. The method combines a self-similar solution \citep{Low1980} in the deep layers  with a current-distributed solution \citep{Pizzo1986} in the upper layers. This class of models allows the freedom to select the values of several parameters which control the properties of the resultant sunspot model. The magnetic configuration is given by the parameters $B_0^L$, $\eta$, and $a$. The former produces changes in the magnetic field strength, while the magnetic field inclination, the curvature of the field lines, and the radius of the structure depend on the other two parameters. The model used as a boundary condition at the axis of the sunspot can be shifted up or down on the axis in order to account for the Wilson depression $z_{\rm WD}$. The last parameter that can be modified is the height $z_0$ where the self-similar and current-distributed solutions are merged. The space of parameters explored includes umbral photospheric field strengths of $B_{\rm u}=$ 1500, 2500, and 3500 G and Wilson depressions of $z_{\rm WD}=$ 250, 350, 450, and 550 km. All the possible combinations of these two parameters were computed, producing a total of twelve independent simulations. The parameters $\eta$, $a$, and $z_0$ are the same for all the models. The radial variation of the photospheric vertical magnetic field approximately follows a Gaussian distribution with a FWHM of 19 Mm (a Gaussian is set at the height $z_0$, but it is modified through an iterative process to concatenate the self-similar and current-distributed models and obtain a MHS solution). The use of models with different magnetic field strength but otherwise the same field geometry allow us to evaluate the effect of field strength on the seismic signal independently of the magnetic field inclination. Figure \ref{fig:cutoffs} shows the vertical stratification of the isothermal cut-off frequency for all the models at the center of the sunspot. As expected, an increase in the Wilson depression increases the depth of the upper turning point. In the following, we will call the ``upper turning point'' the depth where the wave frequency is equal to the local cut-off frequency, that is, to the upper turning point of the radial mode. The depth of the $\beta=c^2/v_{\rm A}^2=1$ layer, where $c$ is the sound speed and $v_{\rm A}$ is the Alfv\'en speed, increases with the magnetic field. This layer is located at a similar depth for all the models with \hbox{$B_u=$3500 G}, but for lower field strengths the depth of the $\beta=1$ layer depends on the Wilson depression. Figure \ref{fig:csonido} illustrates the sound speed. Models with the same Wilson depression but different field strengths present differences in their sound speed, since their gas pressure stratifications differ due to the pressure deficit imposed by the magnetic field. 

A quiet Sun simulation with the same set up of the sunspot cases was performed for applying the method of noise subtraction \citep{Werne+etal2004}. Waves are excited by sources introduced in the equations. They are located at \hbox{150 km} below the photosphere in quiet-Sun regions and their depth increases toward the axis of the spot following a constant temperature surface from the simulation with \hbox{$B_u=$ 2500 G} and \hbox{$z_{\rm WD}=$ 450 km}. Wave sources are randomly distributed in the horizontal directions, but all the simulations use the same random distribution. As a result, the same quiet-Sun simulation can be used for applying noise subtraction to all the sunspot cases. The duration of the simulations with \hbox{$B_{\rm u}=1500$ G} and \hbox{$B_{\rm u}=2500$ G} is 8 h, while the simulations with \hbox{$B_{\rm u}=3500$ G} span 7 h of solar time.

Travel times are computed by applying the procedures for surface-focused helioseismic holography \citep[see][]{Braun+Birch2008} to the vertical velocity maps at constant optical depth $\tau=0.01$. Filtering has been applied in $k-\omega$ space following many previous studies \citep[\eg,][]{Duvall+etal1997, Kosovichev+etal2000, Zhao+etal2001, Braun+Birch2006, Birch+etal2009}. We use a set of phase-speed filters described in \citet{Couvidat+etal2006}. Their mean phase speed spans from \hbox{14.87 km s$^{-1}$} (TD2) to \hbox{35.46 km s$^{-1}$} (TD5). A specific pupil function is associated with each of the filters, with larger pupils for higher phase speeds. See \citet{Couvidat+etal2006} for details of the form of the filters and pupils. The wavefield has also been filtered in temporal frequency, isolating bandpasses with widths of 0.5 mHz centered at 2.75, 3.25, 3.75, 4.25, 4.75, and 5.25 mHz. Finally, the travel-time shifts are obtained by subtracting the quiet-Sun travel time from the sunspot travel time. The quiet-Sun travel time was obtained from the analysis of 7 or 8 h temporal series as reference for the \hbox{$B_{\rm u}=3500$ G} cases or lower field strength cases, respectively.

Figure \ref{fig:power_spectra} illustrates the power spectra of the quiet-Sun simulation, with the model S \citep{Christensen-Dalsgaard+etal1996} eigenfrequencies of the $f$ mode and $p_1$ to $p_4$ modes overplotted as a reference. The regions in the $k-\omega$ domain spanned by the frequency and phase-speed filters are indicated.

\begin{figure}[!ht] 
 \centering
 \includegraphics[width=9cm]{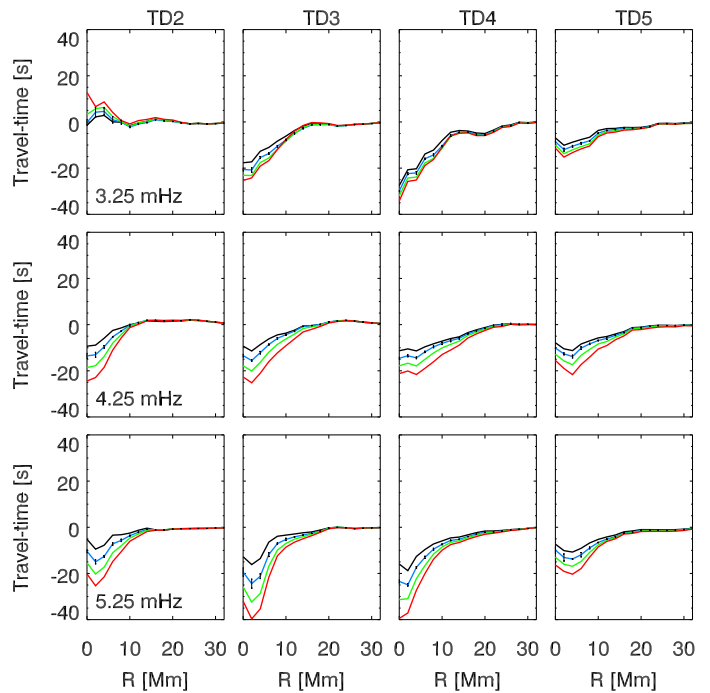}
  \caption{Radial variation of the azimuthally averaged travel-time shifts measured for the simulations with $B_{\rm u}=1500$ G and $z_{\rm WD}=250$ km (black), $z_{\rm WD}=350$ km (blue), $z_{\rm WD}=450$ km (green), or $z_{\rm WD}=550$ km (red). The columns correspond to different phase-speed filters, from lower phase speed (left panels) to higher phase speed (right panels). The rows correspond to frequency filters centered at 3.25 mHz (top panels), 4.25 mHz (middle panels), and 5.25 mHz (bottom panels). The errors bars shown represent an assessment of measurement errors of the travel-time shift differences between two models (see text).}
  \label{fig:radial B1500G}
\end{figure}

\begin{figure}[!ht] 
 \centering
 \includegraphics[width=9cm]{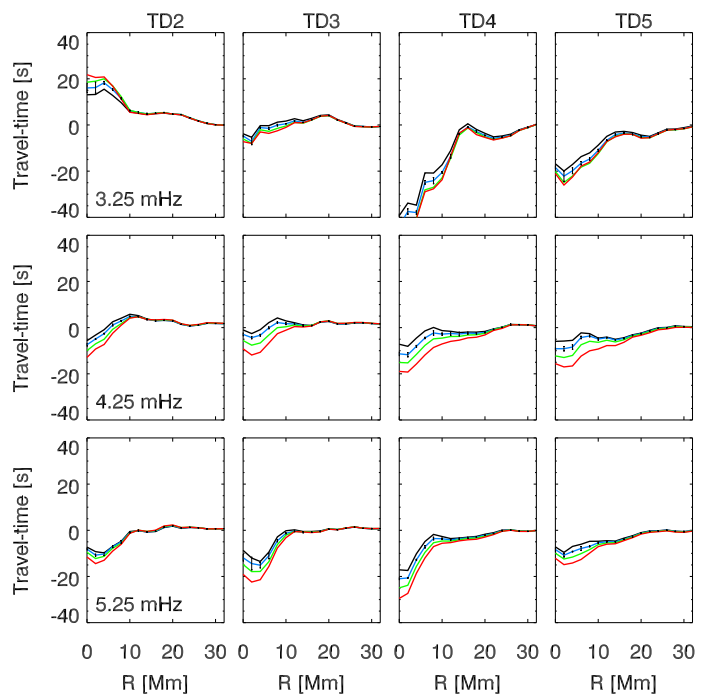}
  \caption{Same as Fig. \ref{fig:radial B1500G} but for sunspot models with \hbox{$B_{\rm u}=3500$ G.}}
  \label{fig:radial B3500G}
\end{figure}

\section{Travel-time shift measurements}
\label{sect:results}

\subsection{Maps}
\label{sect:maps}

The dependence of the travel-time shifts with frequency and phase speed (positive values for low frequencies and low phase speeds and negative values for high frequencies and high phase speeds) is in qualitative agreement with the results obtained from actual observations \citep{Braun+Birch2008, Braun+etal2012} and previous numerical simulations \citep{Moradi+etal2009, Braun+etal2012, Felipe+etal2016a}. Figures \ref{fig:maps B1500G WD550km}-\ref{fig:maps B3500G WD550km} illustrate the travel-time shift maps of the three simulations with a Wilson depression of \hbox{550 km}. We do not show the maps for the other sunspot cases, since their qualitative behavior is similar and the quantitative differences are better appreciated in the radial-average plots in the next sections.

\subsubsection{TD4 and TD5 phase-speed filters}

Negative travel-time shifts are obtained for the phase-speed filters TD4 and TD5 for all frequencies in all of our sunspot models. The magnitude depends on the phase speed and frequency. In the case of the sunspot with \hbox{$B_{\rm u}=3500$ G} (\hbox{Fig. \ref{fig:maps B3500G WD550km}}), the maximum travel-time perturbation is found for the phase-speed filter TD4 and frequency of 3.25 mHz. Higher frequencies show a lower magnitude in the travel time perturbation. On the contrary, the sunspot with the lowest magnetic field strength (Fig. \ref{fig:maps B1500G WD550km}) shows stronger perturbations for high frequencies. For those filter combinations (high phase speed and high frequency) the travel-time perturbation is reduced in amplitude with increasing magnetic field (considering the same Wilson depression).

\subsubsection{TD2 and TD3 phase-speed filters}

The travel-time shifts for low phase-speed filters (TD2 and TD3) changes from positive values (waves apparently slower in the sunspot than in the quiet Sun) to negative values (faster waves in the sunspot) as the frequency increases. The frequency at which this sign reversal is produced is lower for the TD3 filter than for the TD2 filter, but it also depends on the sunspot model. For the sunspot with \hbox{$B_{\rm u}=3500$ G} and \hbox{$z_{\rm WD}=550$ km} (Fig. \ref{fig:maps B3500G WD550km}) the travel-time shift measured at 2.75 mHz and TD3 is mainly positive, while a 3.25 mHz frequency shows a mix of positive and negative shifts. In the other two sunspots with lower field strengths (Figs. \ref{fig:maps B1500G WD550km} and \ref{fig:maps B2500G WD550km}) this combination of filters produces a negative travel-time shift, which is specially prominent in the case with \hbox{$B_{\rm u}=1500$ G}. For the TD3 case, higher frequencies are clearly negative in the three sunspots shown in Figs. \ref{fig:maps B1500G WD550km}-\ref{fig:maps B3500G WD550km}. With regards to TD2 phase-speed filter, in the sunspot with stronger magnetic field negative travel-time shifts are found for frequencies higher than 4 mHz. For the sunspots with \hbox{$B_{\rm u}=1500,2500$ G}, the travel-time shift of waves with 3.75 mHz frequency and phase speed given by TD2 filter shows a negative value surrounded by an annular region with positive shift. For higher frequencies the positive ring vanishes, and only the negative signal remains.

\begin{figure*}[!ht] 
 \centering
 \includegraphics[width=18cm]{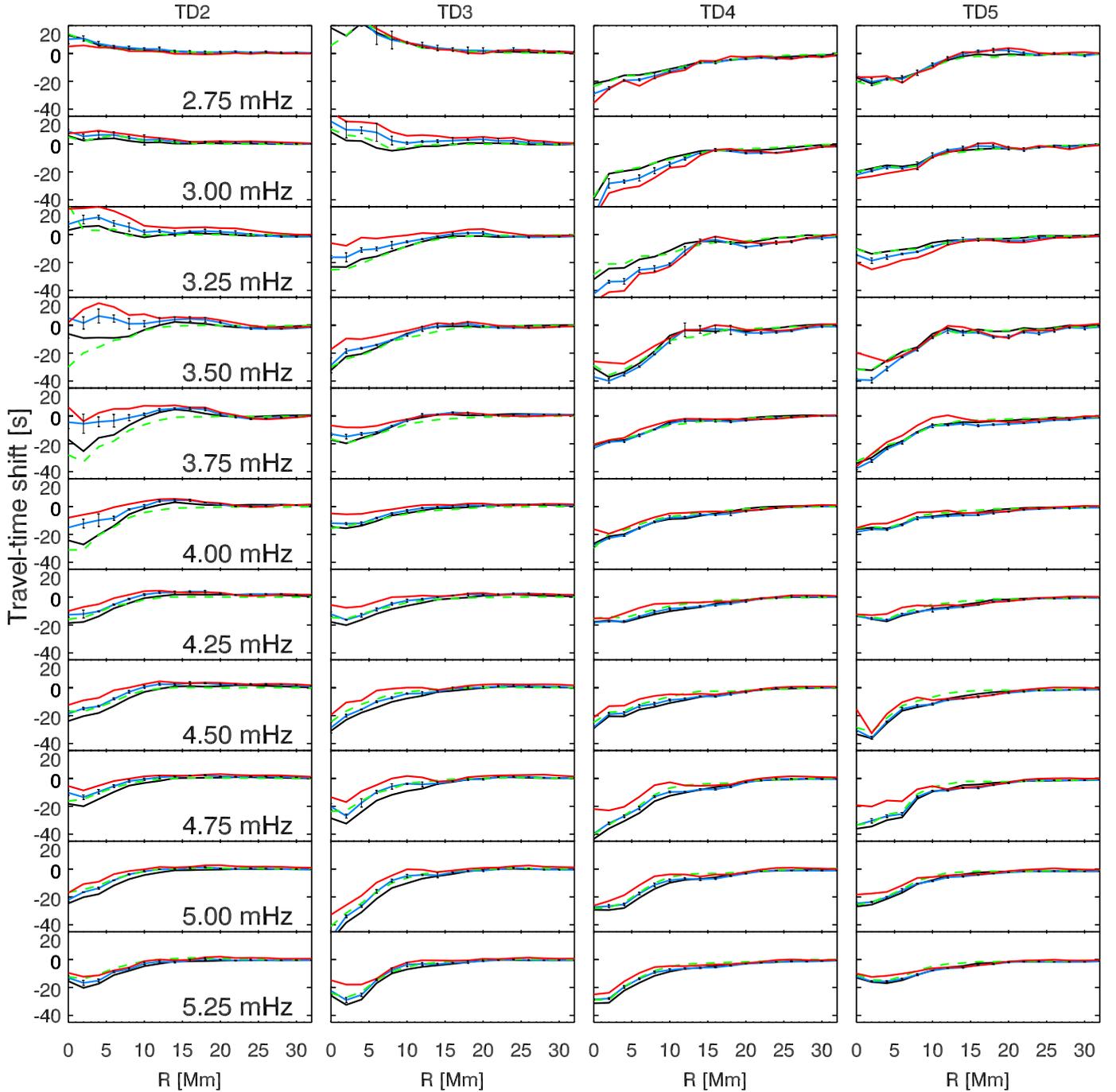}
  \caption{Radial variation of the azimuthally averaged travel-time shifts measured for the simulations with $z_{\rm WD}=450$ km and $B_{\rm u}=1500$ G (black), $B_{\rm u}=2500$ G (blue), or $B_{\rm u}=3500$ G (red). The green dashed line corresponds to the results for a sunspot with the thermal structure of the model with $B_{\rm u}=2500$ G and $z_{\rm WD}=450$ km but with the magnetic field set to zero \citep[see][]{Felipe+etal2016a}. The columns correspond to different phase-speed filters, from lower phase speed (left panels) to higher phase speed (right panels). The rows correspond to frequency filters with increasing frequency from top to bottom. The errors bars shown represent an assessment of measurement errors of the travel-time shift differences between two models (see text).}
  \label{fig:radial WD450G}
\end{figure*}

\subsection{Sensitivity to the Wilson depression}
\label{sect:WD}
We computed azimuthally averaged travel-time shifts for all of the sunspot models. Figure \ref{fig:radial B1500G} shows the radial variation of the averaged travel-time shifts for some selected frequencies for all the simulations with \hbox{$B_{\rm u}=1500$ G}. To assess the significance of differences between measurements of the travel-time shifts
among the models, we compute errors in the following manner. We first take the difference
between the travel-time shift maps for two models with different values of the Wilson
depression. Although the same distribution of acoustic sources are used to reduce the realization
noise (see Sect. \ref{sect:simulations}), a residual amount of noise remains. The map of the travel-time shift difference
is divided into four quadrants centered on the sunspot and the error is estimated as the total spread
(maximum minus minimum value) of the four azimuthal averages within each individual quadrant
(all of the models considered here are cylindrically symmetric, so any variations between
quadrants is due to noise). In order to simplify the figure, only the errors of the difference between the
$z_{\rm WD}=350$ km and 450 km cases are shown, superimposed on the $z_{\rm WD}=350$  km measurements.
Errors computed using other pairs of models are similar. As expected, the measured signal increases in amplitude with the Wilson depression. An increase of the Wilson depression produces a shift in the depth of the upper turning point of the waves toward deeper layers. As a result, the path traveled by wave rays hitting the center of the sunspot is shorter than that of waves reaching the photosphere in quiet-Sun regions. This effect leads to shorter travel times in the sunspot, and the magnitude of this reduction is stronger for more depressed sunspot atmospheres.

\subsubsection{Frequency dependence}

Most of the averaged travel-time shifts illustrated in \hbox{Fig. \ref{fig:radial B1500G}} show a negative travel-time shift. Only the case with TD2 and 3.25 mHz frequency (and lower frequencies with phase speed given by TD2 and TD3, as seen in Fig. \ref{fig:maps B1500G WD550km}) produces a positive perturbation in the travel time. It is interesting to note that this positive perturbation increases with the Wilson depression. In this case, the variation of the travel time cannot be understood as the result of the changes in the depth of the upper turning point. 

The sensitivity of the travel-time shift to the Wilson depression depends on the combination of filters used for the measurement. The travel-time shift measured for waves with 3.25 mHz frequency at the central part of the umbra shows differences around 7 s at most between the results obtained for the case with \hbox{$z_{\rm WD}=250$ km} and the case with \hbox{$z_{\rm WD}=550$ km}. On the contrary, for waves with 5.25 mHz frequency this variation can be larger than 25 s, as found for the phase-speed filter TD3. The top panel of Fig. \ref{fig:cutoffs} shows the vertical stratification of the cut-off frequency from the \hbox{$B_{\rm u}=1500$ G} cases, whose travel-time shifts are plotted in Fig. \ref{fig:radial B1500G}. A wave with 3.25 mHz frequency (bottom horizontal dashed line) propagating upward from the interior will reach the turning point of the \hbox{$z_{\rm WD}=550$ km} model (red line) earlier than that of the smaller Wilson depression models. Waves propagating in sunspot models with small Wilson depression will travel a longer path, closer to that traveled in a quiet-Sun region (thick black dashed line). The difference in the depth where the cut-off frequency is 3.25 mHz between the \hbox{$z_{\rm WD}=250$ km} and \hbox{$z_{\rm WD}=550$ km} cases is around 220 km, in the sense that the path of the later is shorter. For waves with 5.25 mHz, the reduction of the travel path is more than 300 km, producing a larger difference in the travel-time perturbation signal. This effect can qualitatively explain the dependence of the travel-time shift with the frequency for models with the same field strength.

The variation of the travel-time signal with the Wilson depression at constant field strength shows a similar trend for the case with \hbox{$B_{\rm u}=3500$ G} (Fig. \ref{fig:radial B3500G}), although filters TD3 and TD4 show a weaker dependence at the high frequencies. The results for the model with \hbox{$B_{\rm u}=2500$ G} (not shown in the figures) are somewhat between the other two cases.

\begin{figure*}[!ht] 
 \centering
 \includegraphics[width=18cm]{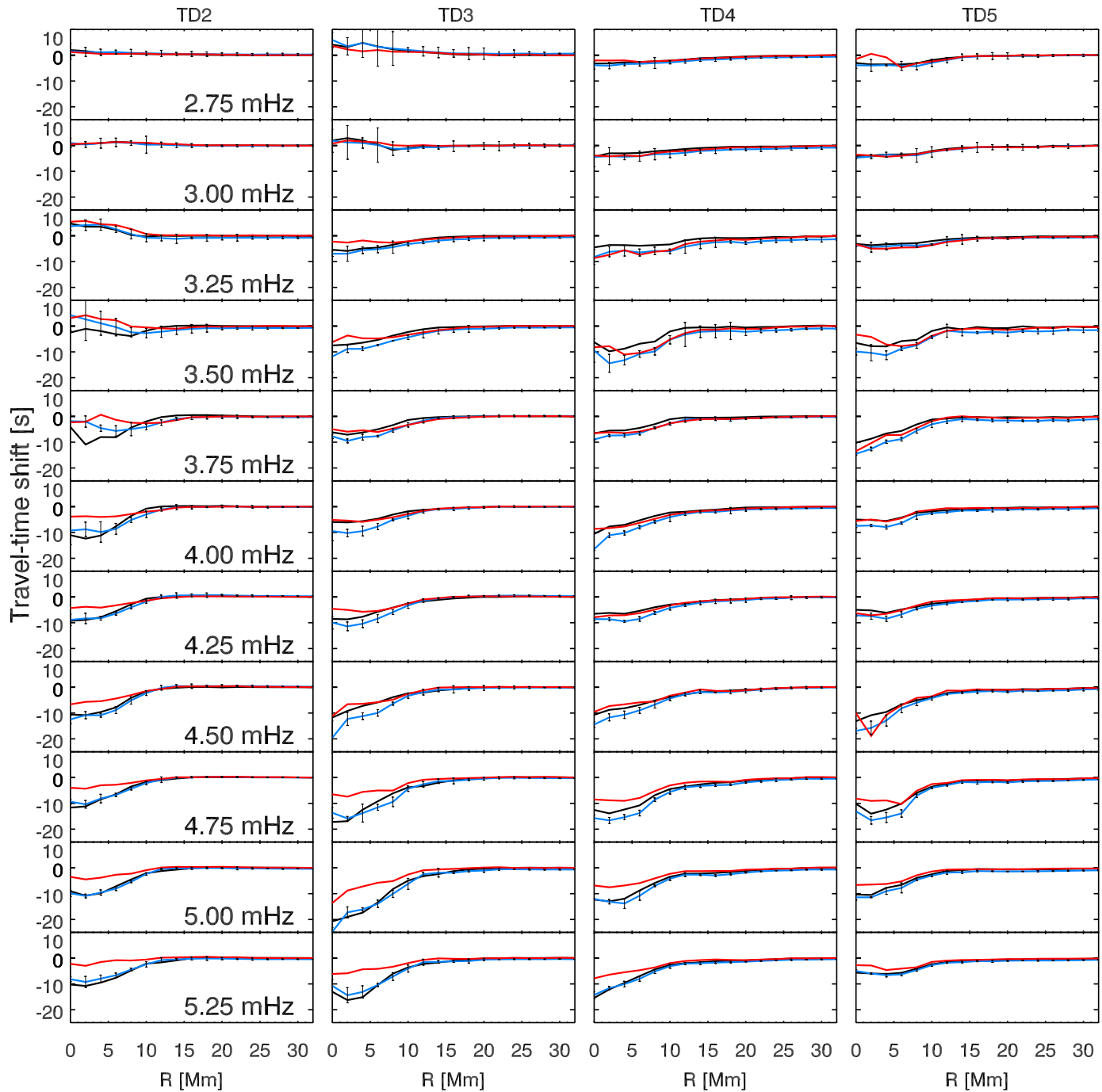}
  \caption{Radial variation of the azimuthally averaged travel-time shifts between models with $z_{\rm WD}=450$ km and models with $z_{\rm WD}=250$ km. The line color indicates the magnetic field strength of the pair of models: $B_{\rm u}=1500$ G (black), $B_{\rm u}=2500$ G (blue), or $B_{\rm u}=3500$ G (red). Negative (positive) differences correspond to shorter (longer) travel times in the model with $z_{\rm WD}=450$ km.  
  }
  \label{fig:radial WD250G}
\end{figure*}

\subsection{Sensitivity to the magnetic field strength}

Figure \ref{fig:radial WD450G} shows the azimuthally averaged travel-time shift measured for the simulations with a Wilson depression of 450 km, but with different magnetic field strengths. Errors are estimated for differences in measurements using models with different magnetic field strength in a manner similar to that discussed in Sect. \ref{sect:WD}. Shown are errors of the difference between measurements from the \hbox{$B_{\rm u}=2500$ G} and the \hbox{$B_{\rm u}=1500$ G} models, superimposed on the \hbox{$B_{\rm u}=2500$ G} measurements. Errors computed using other pairs of models are similar. The largest differences between models are found for the TD2 filter and frequencies between 3.25 and 4.00 mHz. In that region the magnetic field causes changes in the travel-time shift of around 30 s, and the perturbations can even show an opposite sign, as seen in Sect. \ref{sect:maps}. For example, TD2 waves with 3.50 mHz frequency show a negative travel-time shift for the sunspot with \hbox{$B_{\rm u}=1500$ G} but a positive shift for the other two cases. The dashed green lines illustrate the signal obtained from a background model with the density, pressure, and coefficient of specific heats from the sunspot with \hbox{$B_{\rm u}=2500$ G} and \hbox{$z_{\rm WD}=450$ km} but with the magnetic field set to zero. These data correspond to the ``thermal sunspot'' analyzed in \citet{Felipe+etal2016a}.

\subsubsection{Low frequencies}

The travel-time shifts at lower frequencies (2.75 mHz for all the phase-speed filters and also 3.00 mHz for TD2) show a weak dependence on the magnetic field strength. For most phase speed filters and radial positions it is below the error of the measurement procedure. The change in the travel-time shift between models with \hbox{$B_{\rm u}=1500$ G} and $3500$ G is only measurable for the phase speed filter TD2 near the center of the sunspots and for TD4. In all the sunspot models, the upper turning point for low frequencies is deeper than the depth where $\beta=1$ (Fig. \ref{fig:cutoffs}). Upward propagating waves with low frequencies reach the turning point before the magnetic effects are relevant. They neither reach the mode conversion layer nor propagate at depths where the fast speed is significantly modified by the increase of the Alfv\'en velocity.

\subsubsection{Intermediate frequencies}

For frequencies between 3.00 and 3.25 mHz, the negative travel-time perturbations measured with the higher phase-speed filters (TD4 and TD5) increase with the strength of the magnetic field. The case with \hbox{$B_{\rm u}=3500$ G} shows shifts around -40 s for the TD4 filter at the central part of the umbra. The magnitude of the shift is lower for the \hbox{$B_{\rm u}=2500$ G} and \hbox{$B_{\rm u}=1500$ G}, but a comparison of the later with the thermal sunspot reveals a perfect match. Interestingly, although the travel-time perturbations for frequencies between 3.00 mHz and 3.75 mHz are sensitive to the field strength, the thermal sunspot produces signals in quantitative agreement with the sunspot with \hbox{$B_{\rm u}=1500$ G} (except for TD2 filter). That is, the travel times are nearly insensitive to changes in the magnetic field for strengths below 1500 G. As can be seen in Fig. \ref{fig:cutoffs}, the distance between the turning point and the $\beta=1$ depth decreases with the magnetic field strength. Waves with frequency around 3.00 mHz propagating through the \hbox{$B_{\rm u}=2500$ G} and \hbox{$B_{\rm u}=3500$ G} models can be affected by the magnetic effects, but in the \hbox{$B_{\rm u}=1500$ G} model they do not reach the region where these effects are significant. The shorter travel times for the strongly magnetized sunspots agrees with the faster propagation velocity in those models, since their fast magnetoacoustic speed is higher than for the low field strength model. This situation differs from the TD2 and TD3 cases with frequencies between 3.25 mHz and 4.00 mHz, where the models with strong magnetic field show longer travel times. In those cases the interaction with the magnetic field is more complex and must be associated with the larger angle of incidence of those waves, since the lower phase speed waves propagate in a shallower cavity.

\subsubsection{High frequencies}

The travel-time shifts measured with all phase-speed filters show a slight variation with magnetic field for high frequencies (generally around 10 s). In these cases the magnitude of the perturbation decreases with the field strength, opposite to the results found for low frequency and high phase speed waves. The differences between the \hbox{$B_{\rm u}=1500$ G} and \hbox{$B_{\rm u}=2500$ G} cases are small but higher than the value of the measurement error. The travel time obtained for the stronger magnetic field simulation is clearly distinguishable from the other two cases. However, the ``thermal sunspot'' shows a quantitative agreement with the sunspot model with \hbox{$B_{\rm u}=2500$ G} and \hbox{$z_{\rm WD}=450$ km}. These two simulations have exactly the same thermal structure, they only differ in the absence of magnetic field in the former. Their comparison implies that a 2500 G field strength change does not produce significant changes in the travel time. Instead, the variations identified between the three models with different field strength must be due to the  differences in their thermal structure, even though they have the same Wilson depression. The construction of the MHS models requires the imposition of force balance. The magnetic force is different in the three models and, thus, their thermal structure must necessarily differ. The three models have the same Wilson depression, but show differences in the stratification of pressure, density and coefficient of specific heats which lead to differences in their sound speed, as shown in Fig. \ref{fig:csonido}. The sound speed in the layers between $z=-1.7$ Mm and $z=-0.7$ Mm decreases with increasing field strength. $p-$mode waves propagating through the \hbox{$B_{\rm u}=1500$ G} sunspot will be faster than waves in the other sunspots, leading to the reduced travel time measured for high frequencies in this model. 

The frequency above which the travel-time perturbations can be interpreted as a result of the changes in the Wilson depression and sound speed rather than magnetic field depend of the phase-speed filter. For the TD2 case the threshold is at 4.25 mHz, for TD3 at 4.00 mHz, and for TD4 and TD5 filters at 3.75 mHz.

Models with other value for the Wilson depression show a similar dependence with the magnetic field strength. The results of these cases are in qualitative agreement with those illustrated in Fig. \ref{fig:radial WD450G}. Figure \ref{fig:radial WD250G} shows the travel-time shift between models with \hbox{$z_{\rm WD}=450$ km} and models with \hbox{$z_{\rm WD}=250$ km}. That is, it is equivalent to Fig. \ref{fig:radial WD450G} except that the travel times of the \hbox{$z_{\rm WD}=250$ km} cases were subtracted instead of the travel times obtained from the quiet Sun simulation. This representation highlights the differences between pairs of sunspot models. A negative travel-time shift indicates that the travel time of the waves in the simulation with \hbox{$z_{\rm WD}=450$ km} is shorter and vice versa. In the cases with small Wilson depression, the dependence of the high frequency travel times with the magnetic field strength is lower, since the difference in the sound speed is smaller than in the \hbox{$z_{\rm WD}=450$ km} cases (compare the first and third panels of \hbox{Fig. \ref{fig:csonido}}).

\section{Discussion and conclusions}
\label{sect:conclusions}

We have performed a parametric study of the sensitivity of the travel-time shifts measured using helioseismic holography to sunspot models with different Wilson depressions and magnetic field strengths. This study is a continuation of the paper \citet{Felipe+etal2016a}. The results confirm those from the previous work and reveal new findings. Our main conclusions are:

\begin{itemize} 
 
\item The Wilson depression has a strong effect on the measured travel-time shifts. The depression of the atmosphere lowers the height of the upper turning point of the waves, and the path they travel beneath the sunspot is shorter than that of waves in quiet-Sun regions. This causes a reduction in the travel time. The magnitude of the reduction increases with the Wilson depression (Figs. \ref{fig:radial B1500G} and \ref{fig:radial B3500G}). Low frequencies are less sensitive to this effect, since the change in the depth of their upper turning point is smaller than for high frequency waves (Fig. \ref{fig:cutoffs}).

\item Waves with frequencies below 3.00 mHz show a weak dependence on the magnetic field strength. Their cut-off height is below the region where magnetic field effects are significant. In the case of the TD4 and TD5 phase-speed filters, they show a negative travel-time perturbation, which is in agreement with the deeper turning point of those waves in the sunspot models with respect to the quiet-Sun atmosphere. This is opposite to the positive shifts obtained for the low phase-speed filters (TD2 and TD3). \citet{Braun+Birch2008} and \citet{Birch+etal2009} have argued that this sign change is a property of the chosen data analysis filters.

\item The direct effects of the magnetic field on the travel time are apparent for some combinations of phase speed and frequency filters. For the phase-speed filter TD2, the travel-time perturbation measured for frequencies between 3.25 and 4.00 mHz is clearly modified by the magnetic field strength. For the filter TD3, waves with frequencies between 3.00 and 3.75 mHz are sensitive to the magnetic field, while the filters TD4 and TD5 show variation of the travel-time shift with magnetic field for frequencies between 3.00 and 3.50 mHz. These conclusions are based on the analysis of the sunspots with a Wilson depression of 450 km (Fig. \ref{fig:radial WD450G}), but the results are similar for the other models.

\item For all phase-speed filters analyzed, the travel-time perturbation at high frequencies is likely caused by the changes in the thermal model. The Wilson depression shortens the path of the waves, reducing the travel times. However, the Wilson depression on its own cannot account for the obtained travel-time shifts. The variation of the sound speed (among models with the same Wilson depression) at depths between $z=-1.7$ Mm and $z=-0.7$ Mm is also measurable, and it is responsible of changes around 10 s in the travel time (see Fig. \ref{fig:radial WD450G}).

\item The travel-time shift is insensitive to the direct effects of photospheric magnetic field strengths below 1500 G. As indicated in the previous bullet point, travel-times at high frequencies are insensitive to the magnetic field. Only low frequencies depend on the field strength, but for those filters the travel-time perturbation measured from the sunspot with $B_{\rm u}=1500$ G shows a perfect match with the ``thermal sunspot'', whose magnetic field vanishes (Fig. \ref{fig:radial WD450G}). For small sunspots and pores, the only contribution to the travel time comes from the indirect effect of the magnetic field on the thermal structure, through changes in the sound speed and Wilson depression.

\end{itemize}

We have used a set of forward calculations to show examples of how travel times depend on the properties of various sunspot models. We emphasise that the model sunspots described in this paper are not intendeded to represent any particular observed sunspot (for example, by matching the observed sunspot radius or radial profile of magnetic field) and it is therefore premature to compare the modeled travel times with observed sunspot travel times. It is, however, important to know which of these travel-time variations are detectable above the observational noise levels.
Figure 9 from \citet{Braun+etal2012} shows the umbral averages (and errors) of the travel-time shifts measured in the sunspot in AR11092 using 24 hr of Helioseismic and Magnetic Imager \citep[HMI;][]{Schou+etal2012} data. The error they retrieved is below 10 s for all the phase-speed and frequency filters, while for some filter combinations (especially for TD4) the measured error is significantly lower (below 5 s). For photospheric umbral magnetic field around or below 2500 G the change in the travel-time perturbation produced by 100 km shifts in the Wilson depression could be detected above the noise level in the TD3 and TD4 filters for frequencies around 5 mHz (Fig. \ref{fig:radial B1500G}). Models with $B_{\rm u}=3500$ G do not show such a strong travel time variation with Wilson depression at high frequencies (Fig. \ref{fig:radial B3500G}), and the precision of the Wilson depression estimation would be lower. For other filter combinations the observational noise is comparable or above the travel-time shift associated to changes in the Wilson depression.

The highest sensitivity of the travel-time shift to the Wilson depression is obtained for the phase-speed filter TD3 at 5.25 mHz. This combination of phase-speed and frequency filters encompasses the ridge of the $p_{\rm 2}$ mode (Fig. \ref{fig:power_spectra}). This is in general agreement with \citet{Schunker+etal2013}, who concluded that $p_{\rm 2}$ waves are a good candidate for constraining the depth of the Wilson depression. We also find that the phase-speed filter TD4 at 5.25 mHz ($p_{\rm 3}$ mode) is another good candidate for measuring the Wilson depression. This mode was not considered by \citet{Schunker+etal2013}.

The travel-time shifts caused mainly by the changes in the sound speed among models with the same Wilson depression but different field strengths (high frequencies, Fig. \ref{fig:radial WD450G}) are also above the observational noise level when the sound speed perturbation is high enough (sound speed changes between models with $B_{\rm u}=3500$ G and $B_{\rm u}=1500$ G). \citet{Schunker+etal2013} evaluated the sensitivity of the travel time of $p_1$ waves to sound-speed perturbations located at $z=-1.5$ Mm (close to the depth of the sound-speed perturbation in our models). They found that the $p_1$ travel-time shift is below the observed noise level, even for cases with a sound speed perturbation much higher than that introduced in our models. Our results are more sensitive to the changes in the sound-speed because they include higher order $p-$modes (according to the phase-speed and frequency filters used in this work, we have measured up to $p_4$). Interestingly, we found that the lowest frequency for which the travel-time shifts depend on the sound-speed perturbation decreases with increasing phase speed, which agrees with the fact that higher phase-speed filters are sensitive to higher order $p-$modes at lower frequencies (see Fig. \ref{fig:power_spectra}). The depth sensitivity of $p_2$ to $p_4$ modes provides a better sampling of the region where the sound-speed perturbation is located in our models.

\citet{Felipe+etal2016a} suggested a path toward simplified travel-time inversion methods by selecting some combinations of phase-speed and frequency filters that are less sensitive to the magnetic field. The inversion should account for the changes in the Wilson depression and sound speed. This approach would eliminate the need to compute the sensitivity of wave travel times to the strength and geometry of the  magnetic field \citep[\eg,][]{Hanasoge+etal2012}. Our results support this suggestion.

\begin{acknowledgements} 

We acknowledge the financial support by the Spanish Ministry of Economy
and Competitiveness (MINECO) through projects AYA2014-55078-P, AYA2014-60476-P and AYA2014-60833-P. D.C.B acknowledges support from the NASA Living With a Star Program through grant NNX14AD42G awarded to NWRA. A.C.B. acknowledges the EU FP7 Collaborative Project ``Exploitation of Space Data for Innovative Helio- and Asteroseismology'' (SPACEINN). This work used the NASA's Pleiades supercomputer at Ames Research Center, Teide High-Performance Computing facilities at Instituto Tecnol\'ogico y de Energ\'ias Renovables (ITER, SA), and MareNostrum supercomputer at Barcelona Supercomputing Center.

\end{acknowledgements}

\bibliographystyle{aa} 
\bibliography{biblio.bib}

\end{document}